\documentstyle{elsart}
\input psfig.tex
\begin{document}

\begin{frontmatter}
\title{Can high energy neutrino annihilation on relic neutrinos generate
  the observed highest energy cosmic-rays?}
\author{E. Waxman}
\address{Institute for Advanced Study, Princeton, NJ 08540; 
 e-mail: waxman@sns.ias.edu}
\begin{abstract}

Annihilation of high energy, $\sim 10^{21}$eV, neutrinos on big bang
relic neutrinos of 
$\sim 1$eV mass, clustered in the Galactic halo or in a nearby galaxy cluster 
halo, has been suggested to generate, through hadronic $Z$ decay, high energy 
nucleons and photons
which may account for the detected flux of $>10^{20}$eV cosmic-rays. 
We show that the flux of high energy nucleons and photons 
produced by this process is 
dominated by annihilation on the uniform, non-clustered, neutrino background, 
and that the 
energy generation rate of $\sim 10^{21}$eV neutrinos required to account for 
the detected flux of $>10^{20}$eV particles is 
$>10^{48}{\rm erg/ Mpc}^3{\rm yr}$. This energy generation rate, comparable to 
the total luminosity of the universe, is $\sim4$ orders of magnitude larger 
than the 
rate of production of high energy nucleons required to account for the flux of 
$>10^{19}$eV cosmic-rays. Thus, in order for neutrino annihilation to 
contribute significantly to the detected flux of $>10^{20}$eV cosmic-rays, the 
existence of a new class of high-energy neutrino sources, likely unrelated to 
the sources of $>10^{19}$eV cosmic-rays, must be invoked. 

\end{abstract}
\end{frontmatter}

\section{Introduction}

The Fly's Eye \cite{Fly} and AGASA \cite{AGASA} experiments confirmed the 
existence of a break in the energy spectrum of 
high energy cosmic rays at $\sim5\times10^{18}{\rm eV}$, 
for which evidence existed with
weaker statistics in the data of previous experiments (Haverah Park,
Yakutsk, Sugar, see e.g. \cite{Watson} for a review). Fly's Eye data also
strengthen the evidence for a change in primary composition from
predominantly heavy nuclei below the break to predominantly light 
nuclei above the break. These features strongly suggest, when
coupled with the lack of anisotropy that would be expected for 
cosmic-rays (CRs) of Galactic origin, that below $\sim
10^{19}{\rm eV}$ the CRs are mostly heavy ions of Galactic origin, 
and that an extra-Galactic component of protons dominates above 
$\sim10^{19}{\rm eV}$. This conclusion is further supported by the fact that 
the CR energy spectrum is consistent with a cosmological
distribution of sources of protons, with injection spectrum
$dN/dE\propto E^{-2.2}$ typically
expected for Fermi acceleration \cite{cosmology}. In particular, there is 
evidence for the existence of a Greisen-Zatsepin-Kuzmin (GZK) 
``cutoff'', i.e. for the suppression of CR 
flux above $\sim5\times10^{19}{\rm eV}$ expected 
due to interaction of protons with the 
microwave background radiation \cite{GZK}. 

The evidence for GZK suppression 
is strengthened by recent AGASA data \cite{AGASA1}. 
In Fig. 1, the CR spectrum
reported by the Fly's Eye and the AGASA experiments \cite{Fly,AGASA1} 
is compared with the flux expected for 
a homogeneous cosmological distribution of sources, each generating
a power law differential spectrum of high energy protons
$dN/dE\propto E^{-2.2}$ (For the model calculation we have used a flat universe
with zero cosmological constant, $H_0=75{\rm km}\ {\rm s}^{-1}$, and 
time independent energy generation rate per comoving unit volume
$5\times10^{44}{\rm erg/Mpc}^3{\rm yr}$; The 
spectrum is insensitive to the cosmological parameters and to source
evolution, since most of the cosmic rays arrive from distances
$<1$Gpc \cite{cosmology}). The deficit in the number of events 
detected above $5\times10^{19}{\rm eV}$, compared to a
power-law extrapolation of the flux at lower energy,
is consistent with that expected due to a cosmological GZK suppression. 
However, with current data the ``cutoff'' is detected with only 
$2\sigma$ significance \cite{Tokyo}. 

The number of events detected above $10^{20}$eV 
is consistent with that expected
based on the cosmological model presented in Fig. 1 (There is an apparent 
``gap'' between the highest and second highest energy events detected by
the Fly's Eye \cite{gap}. However, assuming that the cosmological model is
valid, the probability that such an apparent ``gap'' would be observed is 
$\sim15\%$ 
\cite{cosmology,gap}). Nevertheless, the detection of $>10^{20}$eV events
does pose challenges to most models of 
CR production. The high energies rule out most of the acceleration
mechanisms so far discussed \cite{Hillas}, and
since the distance traveled by such particles must be smaller
than $100{\rm Mpc}$ \cite{dist} due to their interaction with the
micro-wave background, their arrival 
directions are inconsistent with the position of astrophysical objects,
e.g. jets of powerful radio galaxies \cite{Biermann},
that are likely to produce high energy particles \cite{obj}.

Cosmological $\gamma$-ray bursts (GRBs) are likely sources
of high-energy CRs, which may account for the CR flux above $10^{19}$eV
as well as for the $>10^{20}$eV events \cite{GRBs}. This model recently gained
support form GRB afterglow observations \cite{AG}.
Other models for the production of ultra-high energy CRs were suggested,
where the highest energy events are produced by the decay of super-massive
elementary particles related to grand unified theories 
(see, e.g., \cite{Berezinsky} for recent review). Sources of such particles
may be topological defects, left over from a phase transition 
associated with the symmetry breaking of the grand unified theory \cite{TD}.
While no firm prediction exists of the CR flux in these theories, a generic 
feature of the super-massive particle decay scenarios is that the
injection spectrum is much harder than expected for Fermi acceleration.
Therefore, this scenario can account only 
for the flux of $>10^{20}$eV particles, and can not simultaneously explain the
origin of $10^{19}$--$10^{20}$eV CRs.

It has recently been suggested that 
annihilation of high energy, $\sim 10^{21}$eV, neutrinos on big bang
relic neutrinos of 
$\sim 1$eV mass, clustered in the Galactic halo or in a nearby galaxy cluster 
halo, may generate high energy nucleons and photons
which may account for the detected flux of $>10^{20}$eV cosmic-rays
\cite{Weiler}. 
The existence of $>10^{21}$eV neutrino flux was argued plausible based on the
argument that the mechanism producing the observed high energy, $>10^{19}$eV,
particles, most likely protons, 
also produces charged pions of comparable energy, which subsequently 
decay to produce neutrinos. It was suggested that the generation spectrum 
extends well beyond $10^{20}$eV, and that while nucleons produced by 
a distant source, e.g. a powerful radio galaxy, 
lose their energy interacting with the micro-wave background, 
high-energy neutrinos propagate without energy losses and may annihilate 
on relic neutrinos, producing $>10^{20}$eV nucleons and photons 
at small distances that would allow them to propagate to Earth.
In Sec. 2 we derive the energy density of high energy neutrinos required
to account for the observed rate of $>10^{20}$eV air showers. The implications
of our results are discussed in Sec. 3.

\section{The high energy neutrino density}

Hot Big Bang cosmology predicts the existence of a neutrino
background, similar to the photon micro-wave background. 
The energy distribution of neutrinos
(of mass $m_\nu$ below 1MeV, the decoupling temperature) is predicted to be
a Fermi-Dirac distribution with temperature smaller than that of the photon
background, $T_\gamma=2.73$K, by a factor $(11/4)^{1/3}$. The 
corresponding number density of neutrinos is $n_B=54\ {\rm cm}^{-3}$,
with a similar density of anti-neutrinos
(we assumed a single helicity state; the presence of two helicity states would
increase the density by at most a factor 2, and will not change the
conclusions derived below). Neutrino mass 
smaller than $\sim5$eV is consistent with Big Bang cosmology
\cite{HDM}. Furthermore, there are several indications that neutrinos are
indeed massive. The simplest explanation for the discrepancy between
observed and predicted solar neutrino fluxes (see \cite{Bahcall} for a review)
involves $\nu_e-\nu_\mu$ oscillations driven by neutrino mass difference,
and the common explanation to the atmospheric neutrino anomaly involves
$\nu_\mu-\nu_\tau$ oscillations. While the mass difference implied 
for $\nu_e-\nu_\mu$ oscillations is much smaller than $1{\rm eV}^2$, the
common explanation to the atmospheric anomaly is oscillation to $\nu_\tau$
with mass $\sim0.03$eV \cite{atmo}.

High energy 
neutrinos may interact with the background neutrinos and annihilate, producing 
$Z$-bosons which immediately decay, with $70\%$ probability for
hadronic decay. With standard weak interaction,
the annihilation cross section is strongly peaked at the resonant neutrino 
energy $\epsilon^R=M_Z^2/2m_\nu=4\times10^{21}(m_\nu/1{\rm eV})^{-1}{\rm eV}$,
and the energy-averaged cross section, defined as 
$\bar\sigma\equiv\int ds\sigma_{\nu\nu}(s)/M_Z^2$ 
where $s$ is the square of the 
center of momentum energy, is $\bar\sigma=4\pi G_F/\sqrt{2}=
4\times10^{-32}$cm \cite{sigma}. 
The relative energy width of the resonance peak (FWHM) is 
$\sim3\%$. $Z$ decay produces on average 2.7 nucleons and 
anti-nucleons, and 30 high energy photons through $\pi^0$ decay.
A single nucleon carries, on average, a fraction $\sim0.025$ of the
$Z$ energy \cite{PDG}, and a single photon carries, on average, energy lower
by a factor of $\sim10$ compared to the nucleon energy \cite{Weiler,PDG}. 
For a neutrino mass $m_\nu<10$eV for the background neutrinos, 
the highest energy nucleons produced by the decay
of $Z$-bosons created from annihilation of resonant ($\epsilon^R$) high energy 
neutrinos and background neutrinos have energy $\ge2\times10^{20}$eV, and
could in principle account for the observed highest energy CR events.

Let us first consider the rate of high energy showers due to annihilation
of high energy neutrinos with a homogeneous neutrino background.
The rate per unit volume of $Z$ production is
\begin{equation}
\dot n_Z=\int d\epsilon {dn\over d\epsilon}\sigma_{\nu\nu}(\epsilon)
cn_B,
\label{nZ}
\end{equation}
where $n_B$ is the number density of background neutrinos (or anti-neutrinos)
and $dn/d\epsilon$ is the number density per unit energy of high energy 
neutrinos. We note here that, as suggested by oscillation evidence, neutrino
flavor eigenstates may be different than their mass eigenstates, and since
neutrinos produced by astrophysical sources are expected to be electron
and muon neutrinos from pion decay, 
the $Z$ production rate is smaller than given by
(\ref{nZ}) by a factor $|<\nu_\alpha|\nu_m>|^2$, where $|\nu_\alpha>$
is the flavor eigenstate and $|\nu_m>$ is the massive neutrino eigenstate.
If, as suggested by present data, the massive, $\sim1$eV, 
neutrino is mainly mixed with $\nu_\tau$, then the annihilation
rate of $\nu_{e,\mu}$ expected to be produced by astrophysical sources
will be smaller than given in (\ref{nZ}). However, since we are interested
in obtaining a lower limit to the density of high energy neutrinos required
to produce $\dot n_Z$ high enough to account for the observed flux of CRs
above $10^{20}$, we conservatively assume that $|<\nu_\alpha|\nu_m>|^2$ is
of order unity for the high energy neutrinos.

The integral in (\ref{nZ}) is dominated by the contribution from
$\epsilon\sim\epsilon^R$. We may therefore 
replace the derivative $dn/d\epsilon$ in the 
integrand with its value at $\epsilon=\epsilon^R$. For $m_\nu\sim1$eV the 
background neutrinos are nonrelativistic, and the square of the center of
momentum energy is $s=2\epsilon m_\nu c^2$.  Changing the integration
variable $\epsilon$ to $s$, we find
\begin{equation}
\dot n_Z=cn_B\bar\sigma\left[\epsilon{dn\over d\epsilon}\right]_
{\epsilon=\epsilon^R}.
\label{nZa}
\end{equation}
The high-energy protons and photons produced by $Z$ decay lose energy as they
propagate through the microwave background. 
This limits the distance out to which $Z$ decays 
contribute to the observed rate of air showers above $10^{20}$eV
to $r_{CR}<100$Mpc. 
The flux per solid angle of high energy cosmic-rays producing
showers above $10^{20}$eV, and originating in $Z$ decays, is 
therefore
\begin{equation}
j_{CR}={1\over4\pi}\dot n_Z N_{CR}(m_\nu) r_{CR},
\label{jCR}
\end{equation}
where $N_{CR}(m_\nu)$ is the average number of protons and photons above
$10^{20}$eV produced by the decay of a $Z$-boson of energy $\epsilon^R$.

Using eqs. (\ref{nZa},\ref{jCR}) we may derive the number density 
$[\epsilon dn/\d\epsilon]_{\epsilon=\epsilon^R}$ of high
energy neutrinos in the energy range $\Delta\epsilon\simeq\epsilon^R$ around
$\epsilon=\epsilon^R$, required to produce
a given CR flux $j_{CR}$,
\begin{equation}
n^R\equiv\left[\epsilon{dn\over d\epsilon}\right]_{\epsilon=\epsilon^R}=
{4\pi j_{CR}\over c n_B\bar\sigma r_{CR} N_{CR}(m_\nu)}.
\label{n}
\end{equation}
The total energy density associated with high energy neutrinos is expected
to be larger by a factor $f_{\Delta\epsilon}>1$ than the energy 
$\epsilon^R n^R$ 
associated with neutrinos in the energy range $\Delta\epsilon\sim\epsilon^R$,
since one does not expect the neutrino energy distribution to be strongly
peaked at $\epsilon^R$. If the high energy neutrino flux is indeed produced
by the sources of $>10^{19}$eV CRs, then one expects the neutrino distribution
to be approximately given by $dn/d\epsilon\propto\epsilon^{-2}$ \cite{UHnu}, 
for which the 
energy density in neutrinos of energy $10^{19}{\rm eV}<\epsilon<\epsilon^R$
is $E_\nu(>10^{19}{\rm eV})=
\epsilon^R n^R\log(\epsilon^R/10^{19}{\rm eV})\simeq5
\epsilon^R n^R$. Thus, the energy density of $>10^{19}$eV neutrinos
required to produce the observed rate of showers above $10^{20}$eV,
$j_{CR}\simeq10^{-16}{\rm m}^{-2}{\rm s}^{-1}{\rm sr}^{-1}$ 
(see Fig. 1), is
\begin{eqnarray}
E_\nu(>10^{19}{\rm eV})=&&f_{\Delta\epsilon}
{4\pi cM_Z^2 j_{CR}\over n_B\bar\sigma r_{CR} m_\nu N_{CR}(m_\nu)}\cr
\simeq&&10^{-16}(f_{\Delta\epsilon}/5)
\left[{N_{CR}(m_\nu)m_\nu\over3{\rm eV}}{r_{CR}\over100{\rm Mpc}}
\right]^{-1}\,{\rm erg/cm}^3.
\label{E}
\end{eqnarray}
$E_\nu$ depends only weakly on $m_\nu$, since $N_{CR}(m_\nu)m_\nu\simeq3$eV
independent of the value of $m_\nu$. For $m_\nu\sim1$eV, the resonant energy
is $\epsilon^R\sim4\times10^{21}$eV, and the average energy of the nucleons 
produced in the $Z$ decay is $\simeq\epsilon^R/40\sim10^{20}$eV. Since
the average number of nucleons produce in the decay is 2.7, we have
$N_{CR}(m_\nu)m_\nu\simeq3$eV for $m_\nu\sim1$eV. For higher $m_\nu$
the resonant energy decreases, and only a small fraction of the resulting
nucleons have energy above $10^{20}$eV. For $m_\nu\sim10$eV, 
$\epsilon^R\sim4\times10^{20}$eV and since (on average) 
only $\sim10\%$ of the nucleons
carry $\sim1/4$ the $Z$ energy \cite{PDG}, we have 
$N_{CR}(m_\nu)m_\nu\sim3$eV for $m_\nu\sim10$eV. For $m_\nu\sim0.1$eV,
the resonant energy is $\epsilon^R\sim4\times10^{22}$eV, so that the average
energy of the photons produced in the decay is $\sim\epsilon^R/400\sim
10^{20}$eV, and the photons also contribute to $j_{CR}$. Since the
average number of photons produced in the decay is 30, we have 
$N_{CR}(m_\nu)m_\nu\simeq3$eV for $m_\nu\sim0.1$eV.

The energy density $E_\nu$, required for $Z$ decays to contribute significantly
to the observed rate of $\ge10^{20}$eV air-showers, may be reduced
if the local density of background neutrinos is higher than the average density
$n_B$, due to clustering. For $m_\nu\sim1$eV, the background
neutrinos are non-relativistic at present, and are expected to cluster in 
the gravitational potential wells of galaxies and clusters. Let us first 
consider clustering in the halo of the Galaxy. The density of neutrinos in
the Galactic halo depends on the details of the halo formation. However,
Pauli's exclusion principle allows to put
an upper limit to the number density $n_C$ of neutrinos
clustered in the Galactic halo. This limit is approximately given by
$n_C\le(4\pi/3)(p_{\rm max}/h)^3$, where
$p_{\rm max}\simeq2^{1/2}m_\nu v_r$ is the maximum momentum of bound neutrinos
and $v_r=220{\rm km/s}$ is the Galactic rotation 
speed. A limit lower by a factor of
2 is obtained by requiring the neutrino phase space density, 
approximately given by $n_C/(4\pi p_{\rm max}^3/3)$, 
to be smaller than the maximum phase space density
of the uniform background, $1/2h^3$. This requirement follows from the 
assumption that neutrino clustering results from gravitational collapse
of the uniform neutrino background. In this process, the phase
space density is conserved along particle trajectories, and therefore
the maximum phase space density can not increase \cite{TnG}. The 
neutrino density in the halo is therefore constrained by
\begin{equation}
n_C<1.5\times10^3\left({m_\nu\over1{\rm eV}}\right)^3
\left({v_r\over220{\rm km/s}}\right)^3\,{\rm cm}^{-3}.
\label{nG}
\end{equation}
Using eqs. (\ref{nZa}) and (\ref{jCR}), 
replacing $n_B$ with $n_C$ and $r_{CR}$ with the halo radius $r_H$, we find
that the ratio of the flux of $>10^{20}$ nucleons and photons due
to $Z$ production by annihilation with neutrinos clustered in the Galactic 
halo to the flux due to annihilation on the uniform background is
\begin{equation}
{f_G/f_B}<0.01
\left({m_\nu\over1{\rm eV}}\right)^3{r_H\over50{\rm kpc}}.
\label{EG}
\end{equation}
Thus, for $m_\nu\sim1$eV,
the rate of $>10^{20}$eV showers due to annihilation on neutrinos clustered
in the Galactic halo is two orders of magnitude smaller than that due to
annihilation on the uniform neutrino background. The contribution of 
Galactic halo neutrino annihilation increases with $m_\nu$, and for 
$m_\nu=10$eV may be larger than the contribution due to annihilation
on the uniform background by a factor $\sim10$.

Finally, let us consider the contribution to
$j_{CR}$ due to annihilation on
background neutrinos clustered in a nearby galaxy cluster. The rate of
$Z$ production in the cluster is
\begin{equation}
\dot N_Z=f_\nu{M_c\over m_\nu}\bar\sigma c\left[\epsilon{dn\over d\epsilon}
\right]_{\epsilon=\epsilon^R},
\label{NZ}
\end{equation}
where $M_c$ is the cluster mass, and $f_\nu$ is the fraction of cluster mass
contributed by (background) neutrinos. The cluster mass is given by its 
velocity dispersion $\sigma_v$ and radius $R$, $M_c=2R\sigma_v^2/G$.
The ratio of high energy shower rate
due to annihilation in the cluster, $\dot N_ZN_{CR}/4\pi d^2$ where $d$ is the
cluster distance, to that due to annihilation on the uniform background,
$\dot n_ZN_{CR}r_{CR}$, is
\begin{equation}
f_C/f_B\simeq0.03\left({m_\nu\over1{\rm eV}}\right)^{-1}
\left({d\over20{\rm Mpc}}\right)^{-2}{f_\nu\over 0.1}{R\over1{\rm Mpc}}
\left({\sigma_v\over600{\rm km/s}}\right)^{2}
\,.
\label{EC}
\end{equation}
Clearly, the contribution to high energy showers from neutrino annihilation in 
a nearby galaxy cluster is much smaller that the contribution from annihilation
on the uniform background. We note here that
in deriving eq. (\ref{NZ}) we have assumed that the neutrinos are 
non-degenerate, so that the fraction $f_\nu$ is not limited by Pauli's
principle. This is not valid for small $m_\nu$.
The cluster density at radius $R$ is $\sigma_v^2/2\pi G R^2$,
implying a neutrino (anti-neutrino) density
\begin{equation}
n_C=2.5\times10^4(f_\nu/0.1)\left({m_\nu\over1{\rm eV}}\right)^{-1}
\left({R\over1{\rm Mpc}}\right)^{-2}
\left({\sigma_v\over600{\rm km/s}}\right)^{2}{\rm cm}^{-3}.
\end{equation}
This density is lower than the upper limit imposed by eq. (\ref{nG}) for
\begin{equation}
m_\nu>0.7\left({f_\nu\over0.1}\right)^{1/4}
\left({R\over1{\rm Mpc}}\right)^{-1/2}
\left({\sigma_v\over600{\rm km/s}}\right)^{-1/4}{\rm eV},
\label{mmin}
\end{equation}
where we have used $v_r=2^{1/2}\sigma_v$. 
For $m_\nu<0.7$eV the neutrino density
is limited by the exclusion principle, and the neutrino fraction of cluster
mass is proportional to $m_\nu^4$. Thus, although for low
$m_\nu$, $m_\nu\sim0.1$eV, photons produced in the $Z$ decay may contribute
to the particle flux above $10^{20}$eV, 
the decrease in $f_\nu$ implies that the ratio $f_C/f_B$ is increasing with
$m_\nu$ for $m_\nu<0.7$eV.

\section{Implications}

We have shown that the energy density of high energy neutrinos required to
generate the observed flux of $>10^{20}$ CRs by producing high energy nucleons
and photons via the decay of $Z$-bosons resulting from the annihilation
of the high energy neutrinos
on a uniform neutrino background is $\sim10^{-16}{\rm erg/cm}^3$, independent
of neutrino mass $m_\nu$ for $0.1{\rm eV}<m_\nu<10{\rm eV}$ (cf. eq. \ref{E};
the required energy density is higher for $m_\nu<0.1$eV, and for
$m_\nu>10$eV the $Z$ decay
products are not energetic enough to account for the observed 
$>2\times10^{20}$eV CRs).
The $>10^{20}$ CR flux produced by annihilation
on background neutrinos clustered in a nearby galaxy cluster is 2 orders of
magnitude smaller than that due to annihilation on the uniform background,
independent of the value of $m_\nu$ (cf. eq. (\ref{EC})). 
For $m_\nu\sim1$eV, the CR flux resulting from annihilation on neutrinos 
clustered in the halo
of our Galaxy is smaller by at least a factor $\sim100$ than that due to 
annihilation on the uniform
background (cf. eq. (\ref{EG})). 
The relative contribution of Galactic halo neutrino annihilation
increases with increasing $m_\nu$, and may become comparable to the uniform 
neutrino background contribution for $m_\nu\sim5$eV. For $m_\nu\sim10$eV, 
annihilation on Galactic halo neutrinos may dominate, giving a CR flux
higher by a factor $\sim10$ compared to the flux due to annihilation
on the uniform background. In this case, the lower limit to 
the energy density of high energy 
neutrinos required to produce the observed flux of $>10^{20}$eV CRs 
is lowered by a factor of $\sim10$ to
$\sim10^{-17}{\rm erg/cm}^3$.

For $m_\nu\sim1$eV, the energy density in high energy neutrinos,
$\sim10^{-16}{\rm erg/cm}^3$, required to
account for the observed flux of $>10^{20}$eV CRs is $\sim4$ 
orders of magnitude higher than the energy density in observed
$>10^{19}$eV cosmic-rays, $4\pi J E/c\simeq10^{-20}{\rm erg/cm}^{-3}$ 
(see Fig. 1). 
It is important to note here that although the life time of high
energy neutrinos is longer than that of high energy protons, 
since $10^{19}$eV protons suffer energy loss due to pair production on the 
microwave background while neutrinos lose energy only due to redshift, 
the life time of a $10^{19}$eV proton,
$\sim3\times10^9$yr, is comparable to the Hubble time, $\sim10^{10}$yr, 
and therefore to the neutrino life time. Therefore, the average (over time 
and volume) energy generation rate of high energy neutrinos required to 
produce the neutrino energy density necessary to account for
the $>10^{20}$eV events is at least $\sim4$ 
orders of magnitude higher than the generation rate 
$\sim5\times10^{44}{\rm erg/Mpc}^3{\rm yr}$ of 
$>10^{19}$eV protons required to account for
the observed $>10^{19}$eV CR flux \cite{cosmology}. 
This implies that in order for neutrino annihilation to 
contribute significantly to the detected flux of $>10^{20}$eV cosmic-rays, the 
existence of a new class of high-energy neutrino sources, likely unrelated to 
the sources of $>10^{19}$eV cosmic-rays, must be invoked. 
Furthermore, the 
energy generation rate of the high energy neutrino sources
must be $\sim10^{49}{\rm erg/Mpc}^3{\rm yr}$, 
comparable to the total photon luminosity of the universe.

\paragraph*{Acknowledgments.} 
I thank J. N. Bahcall for helpful discussions. This research was partially
supported by a W. M. Keck Foundation grant and NSF grant PHY95-13835.

\newpage
\begin{figure}[t]
\centerline{\psfig{figure=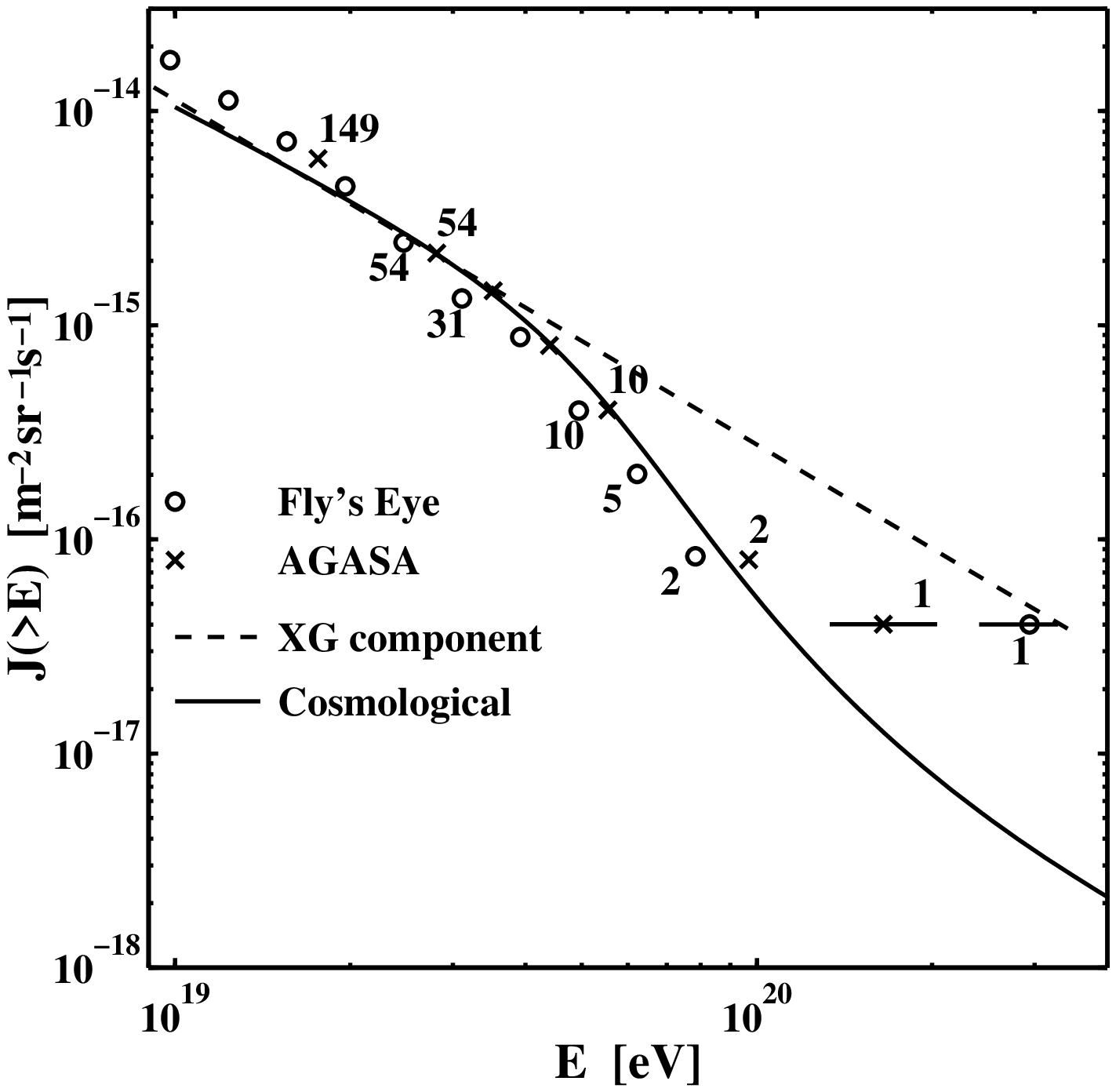,width=6.in}}
\caption{
The CR flux expected for a homogeneous cosmological distribution of sources, 
each generating a power law differential spectrum of high energy protons
$dN/dE\propto E^{-2.2}$, compared to the Fly's Eye 
and AGASA data (The AGASA flux at $3-10\times10^{18}{\rm eV}$
is $\sim1.7$ times higher than that reported by the Fly's Eye, corresponding
to a systematic $\sim20\%$ larger estimate of event energies in the AGASA
experiment compared to the Fly's Eye experiment \cite{Fly,AGASA}; 
We have therefore shifted upward (downward) the Fly's Eye (AGASA) event 
energies by $10\%$). Integers indicate the number of events observed.
$1\sigma$ energy error bars are shown for the highest energy events. 
The dashed line denotes the power law fit by
Bird {\it et al.} \cite{Fly} for the extra-galactic flux dominating above
$\sim5\times10^{18}$eV.}
\label{fig1}
\end{figure}

\end{document}